\documentstyle[12pt]{article}
\begin{document}
\large

\begin{center} {\large \bf 

An integral version of Shor's factoring algorithm} \end{center}

\begin{center} Felix M. Lev\end{center}

\begin{center} Artwork Conversion Software Inc. \end{center}

\begin{center} 1201 Morningside Dr., Manhattan Beach, CA 90266 USA\end{center}

\begin{center} Email: felix@acssouth.com \end{center}

\begin{sloppypar}

\begin{abstract}

We consider a version of Shor's quantum factoring algorithm 
such that the quantum Fourier transform is replaced by an 
extremely simple one where decomposition coefficients
take only the values of $1,i,-1,-i$. In numerous calculations 
which have been carried out so far, our algorithm has been 
surprisingly stable and never failed. There are numerical 
indications that the probability of period finding given by 
the algorithm is a slowly decreasing function of the number 
to be factorized and is typically less than in Shor's 
algorithm. On the other hand, quantum computer (QC), capable 
of implementing our algorithm, will require a much less amount 
of resources and will be much less error-sensitive than 
standard QC. We also propose a modification of Coppersmith'
Approximate Fast Fourier Transform. The numerical  results
show that the probability is signifacantly amplified even in
the first post integral approximation. Our algorithm can be 
very useful at early stages of development of quantum computer.

\end{abstract}

\end{sloppypar}

\section{Motivation}

The discovery of Shor's quantum algorithm for factoring a big 
integer resulted in considerable increase of interest to quantum 
computations and quantum theory in general. The details of the 
algorithm can be found in original Shor's publications [1-3], 
numerous review articles and lecture notes 
(see e.g. Refs. [4-8]). 
The main result of the algorithm is that for quantum computer, 
the number of steps required for factoring a big number ${\bf N}$ 
into primes is of order $(log{\bf N})^3$. It is known that 
even the best classical algorithm requires at least 
$(const{\bf N})^{1/3}$ steps for that purpose. Here and henceforth 
we assume that all logarithms have the base 2. 

The main ingredient of Shor's algorithm is Quantum Fourier 
Transform (QFT - don't confuse this abbreviation with
quantum field theory!). It is used for finding a period of
the function $a^x\,\,(mod\,\,{\bf N})$ where $a$ is a number
coprime to ${\bf N}$. Shor has proved [1-3] 
that the probability of period finding by using QFT is 
asymptotically 
constant when the number is big. However a straightforward 
implementation of the QFT requires computations with 
exponential precision. Therefore it is reasonable to expect 
that any realistic implementation of Shor's algorithm will 
require approximations. Coppersmith \cite{Coppersmith}
has proposed an 
approximate version of the QFT which he called Approximate 
Quantum Fourier Transform (AQFT). In this approach all the 
exponents in question are computed with some accuracy 
and therefore exponential precision can be avoided. The 
problem arises whether such an approximation is stable and 
whether it still guarantees that the probability of period
finding is asymptotically constant. There are indications
\cite{Barenco} that actually the probability is a 
slowly decreasing function of the number to be factorized.

    It is expected that quantum computer (QC) outperforming
classical one (at least for some class of problems \cite{Shor4}) 
will be available in several decades. In all the 
implementations of QC proposed so far, it will require  
substantial overhead resourses in comparison with 
classical computer, and this is believed to be unavoidable 
in view of the nature of quantum theory. It is also believed that
QC will be rather error-sensitive. The problem arises 
whether it is possible, at least at early stages, to 
implement {\bf a real quantum computer}, which satisfies two 
requirements: 

\begin{sloppypar}
\begin{itemize}

\item It will be able to work with numbers, which are rather 
big (say of order $2^{100}$ ) although possibly not so big as 
desired.

\item As compared to classical computer, it will not require
big overhead resources and will have the same order
of (in)sensitivity to errors. 

\end{itemize}
\end{sloppypar}

We believe this problem is solvable and our motivation is 
given below.     

The main difference between classical and quantum computer
is as follows. While each bit in classical computer can have 
only two possible states, which we can denote as $|0\rangle$ 
and $|1\rangle$, 
quantum computer operates with qubits which are quantum 
superpositions of states $|0\rangle$ and $|1\rangle$. This means that 
(at least in principle) each qubit can be prepared in a state 
$c_0|0\rangle + c_1|1\rangle$  where $c_0$ and $c_1$ are arbitrary  
complex numbers. This property of quantum computer (which is 
often called quantum parallelism) makes quantum algorithms 
much more efficient than the corresponding classical ones.
On the other hand, this is just the reason why 
quantum computer requires big overhead resources (in 
comparison with classical one) and is rather 
error-sensitive. The QFT, which is a quantum 
version of fast Fourier transform (FFT) is much more 
efficient because it operates with states $c_0|0\rangle + 
c_1|1\rangle$ where $c_0=1$ but $c_1$ contains phase 
factors $exp(i\alpha)$ with different values of $\alpha$ 
belonging to the field of real numbers $R$ (see below).

The problem of factoring a big integer into primes is 
formulated exclusively in terms of natural numbers, and 
moreover, only a finite range of such numbers is 
involved. One might wonder whether for solving this problem 
it is necessary to involve analytical methods which are 
essentially based on the field of real numbers $R$. In number 
theory there are many examples when propositions related 
entirely to the natural numbers have been proved by using 
powerful analytical methods. On the other hand, many number 
theorists believe (see e.g. Ref. \cite{Davenport}) 
that such propositions 
"should be provable without the intervention of such foreign 
ideas". The history of number theory also contains many 
examples when a proposition related to the natural numbers 
was first proved by analytical methods but then a proof based 
exclusively on natural numbers has been found. 

The above remarks make it reasonable to wonder whether it is 
possible to find a quantum factoring algorithm which involves 
only a finite number of integers. Since quantum computer 
necessarily operates with superpositions of states 
$|0\rangle$ and $|1\rangle$, the problem arises whether quantum 
factoring can be efficient if only combinations 
$c_0|0\rangle + c_1|1\rangle$ with a finite number of integers 
$c_0$ and $c_1$
are involved. Strictly speaking we should make precise the 
following. The power of quantum mechanics is essentially based 
on the fact that the decomposition coefficients can be not 
only real but also complex numbers. Therefore it seems to be 
unwise not to use this power. However we can try to find a 
solution where the coefficients are represented as $c = a + ib$
with only a finite number of integers $a$ and $b$.

Our belief that such a solution can be found, is based on our 
previous investigations of quantum theories where state vectors 
belong not to conventional Hilbert spaces but to spaces over a 
Galois field. As shown in our papers \cite{lev}, it is possible to 
construct a fully discrete and finite quantum theory over a 
Galois field, such that if the characteristic of the field is 
very big then the theory is experimentally indistinguishable 
from the conventional theory based on the field of complex 
numbers $C$. Let us note that Galois fields contain only finite 
numbers of elements which can be treated as positive and 
negative simultaneously. For example, the simplest Galois 
field of characteristic $p$ contains only $p$ elements 
$0, 1, 2,...p-1$. Here $p-1$ plays the role of $-1$, $p-2$ 
plays the role of $-2$ etc. 

The present paper is not based on the results \cite{lev} and we 
assume that the behavior of quantum computer is governed by 
conventional quantum mechanics. Nevertheless, for better 
understanding our motivation for quantum factoring algorithm, 
we describe below our motivation for investigations in \cite{lev}. 

It is quite reasonable to believe that the existing 
mathematics will be insufficient to describe future physics. 
Suppose, for example, that we want to verify experimentally 
whether addition is commutative: $a + b = b + a$. If our 
Universe is finite and contains not more than $N$ elementary 
particles then we shall not be able to do this if $a + b > N$. 
Also it is not clear whether conventional division can be 
always consistent. We know from everyday experience that any 
macroscopic object can be divided by two, three and even a 
million parts. But is it possible to divide by, say, two or 
three the electron or neutrino? We can divide the 
gram-molecule of water by ten, million, billion, but when we 
begin to divide by numbers greater than the Avogadro number 
$6\times10^{23}$, the division operation loses its sense.

A possible objection against quantum theory based entirely on 
integers is that such a fundamental notion as probability 
necessarily involves fractions. In our opinion, the notion of 
probability is a good example for the well-known Kronecker's 
expression that the natural numbers were invented by the God 
and all others were invented by people. Indeed, the notion of 
probability arises as follows. Suppose that conducting 
experiment $n$ times we have seen the first event $n_1$ times, 
the second event $n_2$ times etc. such that $n_1 + n_2 + ... = n$. 
We introduce the quantities $w_i(n) = n_i/n$ (these quantities 
depend on $n$) and $w_i = lim\, w_i(n)$ when $n \rightarrow 
\infty$. Then 
$w_i$ is called the probability of the $ith$ event. We see that 
all information about the experiment under consideration is 
given by a set of integers. However, in order to define 
probability, people introduce additionally the notion of 
rational numbers and the notion of limit. Of course, we can 
use conventional probability even if quantum theory is based 
entirely on integers, but by doing so we should realize that 
it is only a convenient (or common?) way to describe the 
measurement outcome.

Another objection closely related to the previous one, is that 
the notion of unitary transformation also necessarily involves 
fractions. However, the requirement that physical transformations 
must be unitary is not necessary. This requirement is based in 
particular on the assumption that the total probability is a 
conserving physical quantity. Meanwhile, the total probability 
does not have any physical meaning, only relative probabilities 
of different outcomes do. Mathematically this is expressed as 
the statement that Hilbert spaces describing quantum systems 
are projective: the elements $\psi$ and $c\psi$ describe the 
same physical state. Therefore it is quite sufficient to 
require unitarity in projective space: the transformation 
should be unitary up to an arbitrary factor.  

Let us stress again that in the present paper we assume that the 
behavior of quantum computer is governed by conventional quantum 
mechanics. At the same time our algorithm remains unchanged for 
a purely discrete and finite version of quantum theory.

The paper is organized as follows. In Sect. 2 we outline the 
QFT in the 
way convenient for the subsequent presentation of our algorithm 
in Sect. 3. Numerical results are described in Sect. 4 and
concluding remarks are given in Sect. 5.

\section{Outline of quantum Fourier transform}

\begin{sloppypar}

Consider quantum computer operating with $n$ qubits. The Hilbert
space describing all possible states of this computer is the 
tensor product of $n$ spaces describing the qubits in question. 
The dimension of this space is equal to $N = 2^n$. The basis of the
space can be chosen in such a way that the basis element $X$ is 
defined by some natural number $x = 0, 1, 2,...N-1$ as follows. If
\begin{equation}
x = x_{n-1} 2^{n-1} + x_{n-2}2^{n-2} + ... x_02^0
\label{1}
\end{equation}
is a binary expansion of $x$, such that each $x_i$ can be either 
0 or 1, then $X$ is represented as a tensor product
\begin{equation}
     X = |x_{n-1}\rangle|x_{n-2}\rangle|x_{n-3}\rangle...|x_0\rangle
\label{2}
\end{equation}
\end{sloppypar}

If $X'$ is another basis vector defined by $x'$ then, as follows from 
Eqs. (\ref{1}) and (\ref{2}), $X$ and $X'$ will be orthogonal 
if $x \neq x'$. 
Indeed, in that case there exists at least one value of $i$ such 
that $x_i \neq x_i'$ and the orthogonality follows from Eq. (\ref{2}).

Let $Y$ be another basis element identified by $y$ in a similar 
way. Then the quantum Fourier transform (QFT) is defined as an 
operator $F$ which acts on $X$ as follows

\begin{equation}
FX=\frac{1}{\sqrt{N}}\sum_{y=0}^{y=N-1}exp(\frac{2i\pi xy}{N}) Y
\label{3}
\end{equation}

We can rewrite this definition as	

\begin{equation}
FX=\frac{1}{\sqrt{N}}\sum_{y_0,y_1,...y_{n-1}} 
|y_{n-1}>|y_{n-2}>....|y_0>exp(\frac{2i\pi xy}{N})
\label{4}
\end{equation}
where the values of $y_i$ can be either 0 or 1. In the exponent 
we can use the binary expansions for $x$ and $y$, and take into 
account the fact that all multiples of $2^n$ do not contribute 
to the result. Then we arrive at 
\begin{eqnarray}
&& FX = \frac{1}{\sqrt{N}} 
\{|0\rangle + exp[\frac{i\pi}{2}(2x_0)]|1\rangle\}\times\nonumber\\ 
&&\{|0\rangle + exp[\frac{i\pi}{2}(2x_1 + x_0)]|1\rangle\}\times \nonumber\\
&&\{|0\rangle + exp[\frac{i\pi}{2}
    (2x_2 + x_1 + \frac{x_0}{2})]|1\rangle\}....\times\nonumber\\
&&\{|0\rangle + exp[\frac{i\pi }{2}(2x_{n-1} + x_{n-2} + 
\frac{x_{n-3}}{2} + ...\nonumber\\
&& \frac{x_0}{2^{n-2}})]|1\rangle\}
\label{5}
\end{eqnarray} 
This expression is written in the form which will be convenient 
in Sect. 3.

   In Shor's algorithm, the QFT is applied to special periodic 
states which can be described as follows. Let $r  << N$ and 
$x(0)  < r$ be some natural numbers. Consider  the numbers 
$x(j) = x(0) + jr$, where $j = 0, 1,... A-1$ and $A$ is such that 
$x(A-1) < N$, $x(A) \geq N$. Let $X(j)$ be the basis element defined 
by $x(j)$ and
\begin{equation}
  X = \frac{1}{\sqrt{A}} \sum_{j=0}^{A-1} X(j)
\label{6}
\end{equation}
Then, as follows from Eqs. (3) and (6), the probability to find 
a state $Y$ in $FX$ is equal to
\begin{equation}
  Prob(y) = \frac{A}{N}\, |\frac{1}{A} \sum_{j=0}^{j=A-1} 
exp(\frac{2i\pi jry}{N}) | ^2  
\label{7}
\end{equation}
where $Y$ is defined by $y$. 

The simplest case is such that $r$ exactly divides $N$ and 
therefore $N = Ar$. Then $Prob(y)$ equals $1/r$ if $y/N = k/r$ 
$(k=0, 1,... r-1)$ 
and equals 0 for other values of $y$. Therefore with the 
probability 1 the result of the measurement of the state 
$FX$ is such that $y/N$ equals $k/r$ for some $k = 0, 1,... A-1$, 
and we have good chances to find the period $r$. As shown by 
Shor [1-3], if we know $r$ then we have good 
chances to find a 
prime divisor of ${\bf N}$ where ${\bf N}$ is the number to be 
factorized.

Let us now consider a general case. It is easy to show that 
there exist at least $r$ values of $y$ satisfying 
\begin{equation}
|\frac{y}{N} - \frac{k}{r}| \leq \frac{1}{2N} 
\label{8}
\end{equation}
For such $y$ we can estimate the sum in Eq. (7) (see e.g. Refs. 
[1-8]) using the property that if $\alpha$ belongs to the interval 
$[0, \pi /2]$ then $|sin \alpha | \leq \alpha$ and 
$|sin \alpha | \geq 2\alpha /\pi$. 
The final result is that at least with the probability 
$4/\pi^2$ the measured 
value of $y$  satisfies Eq. (\ref{8}) with some $k = 0,1,...r-1$. 
(As shown in Ref. \cite{Ekert2}, the probability can be 
amplified but for that purpose the number of bits (and the
complexity of computations) should be increased).
The values of $k$ and $r$ can be efficiently extracted from the 
values of $y$ and $N$ by the continued fraction method, and we can 
repeat the measurements if necessary. Therefore we have good 
chances to find $r$ and then factorize the number in question.

\begin{sloppypar}
\section{An integral version of quantum Fourier transform}
\end{sloppypar}
It is clear from Eq. (5), that if Shor's algorithm is used in a 
straightforward way then the following problem arises. For big 
values of $N$ (and only such values are of interest) this 
expression contains very small exponents and 
therefore the quantum measurement preparing the state (5) 
should be performed with exponential accuracy. 

\begin{sloppypar}
Coppersmith \cite{Coppersmith} has proposed an approximate 
quantum Fourier 
transform (AQFT) such that all the terms containing $1/2^l$ in 
the exponents are neglected if $l$ is greater than some 
number $m$. Then for each $\epsilon$ specifying the accuracy of the 
exponents, we can find the required value of $m$. The 
complexity of the quantum circuit implementing AQFT becomes 
$O(n\,logn)$ instead of $O(n^2)$ for the QFT. Actually the complexity 
is rather sensitive to the required accuracy. If it is small 
then the complexity is close to $O(n)$ and in the opposite case 
it is close to $O(n^2)$. The problem arises whether the QFT is 
stable under small perturbations of the exponents and 
whether the probability of period finding is reasonably 
high for realistic values of the numbers to be factorized. 
\end{sloppypar}

This problem has been investigated in Ref. \cite{Barenco} and
there exists a vast literature devoted to the role of error
corrections in Shor's algorithm. If $m$ 
is the maximum number of terms retained in each exponent index, 
then, as shown in Ref. \cite{Barenco}, Shor's lower bound 
$4/\pi^2$ should be replaced by 
\begin{equation}
MinProb = \frac{8}{\pi^2} sin^2(\frac{1}{2}(\frac{\pi}{2} - 
\Delta_{max}))
\label{9}
\end{equation}
In the most favorable case $\Delta_{max} = 0$ and we again 
arrive at Shor's result. However such a favorable scenario 
cannot be guaranteed. If $m > log\,n +2$ then it can be 
guaranteed that 
\begin{equation}
     MinProb = \frac{8}{\pi^2} sin^2(\frac{\pi m}{4n}) 
\label{10}
\end{equation}
Therefore in the worst scenario the probability will be a 
function decreasing asymptotically as $(log\,n/n)^2$. 

As discussed in Sect. 1, our goal is to find an algorithm 
which can be formulated exclusively in terms of integers. 
Let us consider how we can modify Eq. (5) to satisfy this 
requirement. First we note that the presence of the 
factor $1/\sqrt{N}$ is irrelevant. This factor is needed 
only to ensure unitarity but, as noted in Sect. 1, it is 
quite sufficient to require that the transformation should 
be unitary up to a constant factor. Let us now consider the 
exponents in Eq. (5). For the most important qubit 
our requirement is satisfied automatically since 
$exp(i\pi x_0)$ can be either +1 or -1. For the second qubit 
this requirement is satisfied too because $exp(i\pi x_0/2)$ can 
be either 0 or $i$ (as noted in Sect. 1, we allow the 
coefficients to be of the form $a + bi$ where $a$ and $b$ are 
integers). For the third qubit the requirement is not always 
satisfied because if $x_0$ equals 1 then the coefficient 
contains $exp(i\pi /4) =(1+i)/\sqrt{2}$. For the subsequent 
qubits the state $|1\rangle$ enters with the coefficient 
$exp(i\alpha )$ where 
\begin{equation}
\alpha = \pi (x_l + \frac{x_{l-1}}{2} + \frac{x_{l-2}}{4} + 
\frac{x_{l-3}}{8} +... \frac{x_0}{2^l})
\label{11}
\end{equation}

We can ensure the integrity of the coefficient by neglecting all 
the terms beginning from $x_{l-2}/4$. In that case we will have the 
the AQFT when the maximum number of terms retained in 
each exponent index equals $m=2$ (the simplest case of the 
AQFT corresponding to m=1 is known as the Hadamard transform). 
The value of $\alpha$ in that case will be always underestimated 
if at least one of the numbers $x_{l-2}, x_{l-3},... x_0$ is not equal 
to zero. In the worst scenario the difference between the exact 
and approximate values of $\alpha$ can be close to $\pi /2$ and
can accumulate for different qubits.   

Another option is to replace the expression (11) by
\begin{equation}
\beta=\pi (x_l + \frac{x_{l-1}}{2} + \frac{x_{l-2}}{2})
\label{12}
\end{equation}
In that case the coefficient in question also will be integral as 
required. If $x_{l-2} = 0$ then $\beta$ is always less or equal 
$\alpha$ and their difference does not exceed $\pi/4$ in the 
worst scenario. On the contrary, if $x_{l-2} = 1$ then $\beta$ is 
always greater than $\alpha$ and in the worst scenario the 
difference also cannot exceed $\pi/4$. One might hope that for
different qubits the both effects may considerably cancel out, 
and the result will be close to that given by the QFT or some 
higher order AQFT. 

A straightforward generalization of our proposal is that for $m>2$
the AQFT in the $m$th approximation can be modified as follows. 
When one considers the contribution of 
\begin{equation}
ind = 2i\pi (x_l + \frac{x_{l-1}}{2} +...\frac{x_0}{2^l})
\label{AFT1}
\end{equation}
to the exponent in the $m$th approximation, then, instead
of retaining $m$ terms as
\begin{equation}
ind_m = 2i\pi (x_l + \frac{x_{l-1}}{2} +...\frac{x_{l-m+1}}{2^{m-1}})
\label{AFT2}
\end{equation}             
we propose also to retain the $(m+1)$th term but with the 
coefficient 2:
\begin{equation}
ind_m' = 2i\pi (x_l + \frac{x_{l-1}}{2} +...\frac{x_{l-m+1}}{2^{m-1}}+
\frac{x_{l-m}}{2^{m-1}})
\label{AFT3}
\end{equation}  	 
The above arguments make it reasonable to think that in that case 
the convergence to the QFT will be better and our results for
$m=3$ in Sect. 4 confirm this. However it is clear that for $m>2$ 
the AQFT is not formulated only in terms of integers.  

To summarize, we are going to investigate the transform which, 
by analogy with Eq. (\ref{5}), reads
\begin{eqnarray} 
&& F_IX = \frac{1}{\sqrt{N}} \{|0\rangle + 
exp[\frac{i\pi}{2}(2x_0)]|1\rangle\}\times\nonumber\\ 
&&\{|0\rangle + exp[\frac{i\pi}{2}(2x_1 + x_0)]|1\rangle\}\times\nonumber\\
&&\{|0\rangle + exp[\frac{i\pi}{2}(2x_2 + x_1 + x_0)]|1\rangle\} ....\times\nonumber\\
&& {|0\rangle + exp[\frac{i\pi}{2}(2x_{n-1} + 
x_{n-2} + x_{n-3})]|1\rangle}
\label{13}
\end{eqnarray}
Here the subscript $I$ in $F$ stands for "integral".

As noted in Sect. 1, the overall normalization factor 
$1/\sqrt{N}$ is irrelevant. We retain it for convenience of the 
readers preferring strict unitarity. It is easy to show that the 
operator $F_I$ is indeed unitary. Indeed, the norm of $F_IX$ is 
equal to 1, i.e. the norm of $X$. Furthermore, if $X$ and $X'$ are 
orthogonal, the same is true for $F_IX$ and $F_IX'$. 
Indeed, as follows from Eq. (\ref{13}), 
the contribution of the leftmost bit to the scalar product 
$(F_IX, F_IX')$ is equal to $1 + (-1)^{x_0 - x_0'}$. This 
quantity is not equal to 
zero only if $x_0 = x_0'$. If this is the case then the 
contribution of the second bit from the left is obviously equal to 
$1 + (-1)^{x_1 - x_1'}$. Analogously, the result is not zero only 
when $x_1 = x_1'$. By repeating this procedure we conclude that 
$(F_IX, F_IX')$ is not equal to zero only if $X=X'$.

Apart from the (irrelevant) normalization factor in Eq. (\ref{13}), 
all the coefficients in front of $|1\rangle$ in this expression 
obviously have only one of four values: $1, i, -1, -i$. 
Therefore the algorithm based on $F_I$ indeed operates only 
with integers.

It is clear from Eq. (\ref{13}) that the quantum circuit 
implementing the transformation $F_I$ has the complexity $O(n)$ 
but the main problem of course is whether our algorithm allows 
to peak the required values of $y$ with a reasonable probability.

We denote
\begin{eqnarray}
&&f(x, y, n) = [2(x_0y_{n-1} + x_1 y_{n-2} +...x_{n-1}y_0)  +\nonumber\\ 
             &&(x_0y_{n-2} + x_1 y_{n-3} +...x_{n-2}y_0)  +\nonumber\\
             &&(x_0y_{n-3} + x_1 y_{n-4} +...x_{n-3}y_0)]\, (mod\,\, 4),\nonumber\\
&&g(x,y, n) = exp[\frac{i\pi}{2} f(x, y,n)]
\label{14}
\end{eqnarray}
The function $g(x, y, n)$ can obviously have only one of the values 
$1, i, -1, -i$.
Then we can rewrite Eq. (\ref{13}) as
\begin{equation}
   F_IX = \frac{1}{\sqrt{N}} \sum_{y=0}^{y=N-1} g(x, y, n) Y
\label{15}
\end{equation}
and apply this transformation to the state defined by Eq. (\ref{6}). 
The overall factorization factors $1/\sqrt{N}$ and $1/\sqrt{A}$ are 
not important, but if we wish to describe different measurement 
outcomes in terms of conventional probability (see the discussion 
in Sect. 1), it is convenient to retain them. As follows from 
Eq. (\ref{15}), the (conventional) probability of the 
measurement outcome $y$ is given by
\begin{equation}
   Prob_I(y) = \frac{A}{N}\, |\frac{1}{A} 
\sum_{j=0}^{j=A-1} g(x(0) + jr, y, n) | ^2
\label{16}
\end{equation}
In the most favorable case, when the quantity $g(x(0) +jr, y, n)$ 
is the same for all the values of $j$, the quantity $Prob_I(y)$ 
is equal to 
$A/N$. As noted in the preceding section, for the QFT this is the 
case, in particular, when $r$ exactly divides $N$ (i.e. $N = Ar$) 
and $y = kA$ for some natural $k$ in the range $0, 1, ...r-1$. 
The same is valid in our case. Indeed, at such conditions we have 
$r = 2^l$ where $l$ is an integer which is much less than $n$ 
(because the algorithm applies only if $r << N$) and $A = 2^{n-l}$. 
The binary expansion of $r$ obviously contains only the $l$th 
nonzero bit. Therefore the first $l-1$ bits in all the numbers 
$x(0) + jr$ are the same. At the same time, the first $n-l-1$ 
bits of the number $y = kA$ are always equal to zero. Therefore, 
as follows from Eq. (\ref{14}), the quantity $f(x, y, n)$ depends only 
on the first $l-1$ bits of $x$ and thus $f(x(0) + jr, y, n)$ is 
indeed the same for all the values of $j$. Let us note that the 
same arguments apply to the QFT (in which case they can be 
treated as a proof based not on geometric series of phase factors
but on 
positions of relevant bits in $x$ and $y$) and to any version of 
the AQFT. Therefore the requirement $Prob_I(y) = A/N$ for such 
conditions does not impose practical restrictions on the algorithm.   

In the general case we did not succeed in finding a good 
estimation for $Prob_I(y)$ and therefore we should perform direct 
numerical computations of this quantity. The results are 
described in the next section.

\section{Numerical results}

It is clear from Eqs. (\ref{14}) and (\ref{15}) that the success of 
straightforward numerical computations of the probabilities in 
question depends mainly on how efficiently the function $g(x, y, n)$
can be calculated. Of course, for particular values of $x$ and $y$ 
this is a trivial task for modern computers. However in real 
computations this function should be calculated for many 
different values of $x$ and $y$, and the time of the computation 
crucially depends on the computational algorithm. As follows from 
Eq. (\ref{14}), the computation of $g(x,y,n)$ requires a direct access 
to each bit and therefore it is reasonable to believe that such 
programming languages as C or C++ (to say nothing about 
assembly language) will be convenient for 
that purpose. Moreover, all the modern implementations of C++ 
compilers include the standard template library (STL) which 
contains a container called bitset. Consider, for example, the 
expression  
\begin{equation}
h(n) = x_0y_{n-1} + x_1y_{n-2} + ... + x_{n-1}y_0
\label{17}
\end{equation}
It is clear that the function $g(x, y, l)$ is defined by $h(l)$ with 
$l = n, n-1, n-2$. Let $z$ be an $n$-bit integer which is a reversal 
of $y$:
\begin{equation}
   z = \sum_{i=0}^{i=n-1} z_i2^i = \sum_{i=0}^{i=n-1} y_{n-i-1}2^i
\label{18}
\end{equation}
Then
\begin{equation}
h(n) = x_0z_0 + x_1z_1 + ... + x_{n-1}z_{n-1}
\label{19}
\end{equation}
We can create two bitsets representing the arrays of bits for $x$ and 
$z$, say $B(x)$ and $B(z)$, respectively. Then it is clear from Eq. 
(\ref{19}), that $h(n)$ is equal to the number of bits in the 
array $B$ 
obtained by ANDing the arrays $B(x)$ and $B(z)$. The STL provides 
the overloaded operator for that purpose (which is called $\&=$) and 
the function count() which says how many nonzero bits the bitset in 
question has. However there are practical inconveniences in using 
this approach. The matter is that in the existing version of the 
STL, the bitset constructor, creating a bitset from the number in 
question, and the inverse function, converting the bitset to a 
number, are implemented only for numbers less than 4294967296 which 
are represented by 32 bits. Therefore for bigger numbers one 
should write his or her own versions of those functions. In any case,
ANDing bits is much faster than a straightforward multiplication 
for computing products in Eq. (\ref{19}).      

The second problem is that for rather small values of 
$N = 2^n$ (say $n \leq 27$ or $N \leq 134217728$) we can compute 
probabilities for all the values of $y$ in a reasonable time. 
However the complexity is growing with $N$ roughly as $N$ and for 
essentially bigger values of $N$ this seems to be unrealistic if 
standard computers are used for that purpose. Let us recall that 
our main goal is to extract the value of $r$ from the measured 
value of $y$. As noted above, for $y$ satisfying Eq. (\ref{8}) 
one can recover the values of $k$ and $r$ by the continued 
fraction method. Are other values of $y$ of any use for us? 

To answer this question we recall how the continued fraction method
is used for extracting the period from the measurement outcome.
If {\bf N} is the number to be factorized then the adopted strategy
is to choose $N= c{\bf N}^2$ where typically
$c$ is a small number (say in the range 2-5) such that $N$ is 
a power of two. The reason is
that on the one hand, for a given {\bf N} we want to work with the
least possible number of bits, but on the other hand we should
unambiguously extract the period $r$. The value of $r$ is always
less than {\bf N} by construction of Shor's algrithm ($r$ is defined
as a period of a function $f(x) = a^x\quad(mod\,{\bf N})$ where
$a$ is a number coprime to {\bf N} [1-3]). 
We use the property of the continued fraction method that if
$k_1/r_1$ and $k_2/r_2$ are two continued fractions for $y/N$
then $k_2/r_2$ approximates $y/N$ better if and only if $r_2 > 
r_1$ (see e.g. Ref. \cite{Davenport}). Therefore our approach 
is as follows.
For a given $y$ we develop continued fractions for $y/N$ and
stop if the next approximation has the denominator greater or
equal than
${\bf N}$. Then for each $y$ we can unambiguously find a
continuos fraction $k_1/r_1$ satisfying the requirement that it
is the best approximation among the continuos fractions with the
denominator less than {\bf N}. However in the general case 
the result may have nothing to do with the period $r$.  

Suppose however that $y$ satisfies
Eq. (\ref{8}) and $k_1/r_1$ is the best approximation obtained in
such a way. If $k_1/r_1 \neq k/r$ then obviously 
$|k/r - k_1/r_1|\leq 1/N$ but on the
other hand this contradicts the obvious fact that they also 
satisfy $|k/r - k_1/r_1|\geq 1/rr_1 > c/N$.
Therefore for $y$ satisfying Eq. (\ref{8}) $k/r$ is the best 
approximation obtained as described above.

Let us now reformulate the problem in this way: if $y$ satisfies 
Eq. (\ref{8}) and $k/r$ is the continued fraction for $y/N$ then can we 
guarantee that for values of $y_1$ close to $y$, $y_1/N$ is 
approximated by the same continued fraction $k/r$?
Suppose that $|y_1-y| = a$, $k_1/r_1$ is the best approximation
for $y_1$ and $k_1/r_1 \neq k/r$. Then on the one hand      	     
$|k/r - k_1/r_1|\geq 1/rr_1 > c/N$ and on the other
$|k/r - k_1/r_1|\leq (2a+1)/N$. This is impossible if $a <(c-1)/2$.
We conclude that, depending on the value of $c$, $k/r$ satisfying
Eq. (\ref{8}) represents also the best approximation for the 
values $y_1$ in some vicinity of $y$.  

Taking into account the above consideration we adopted the
following approach. For $n\leq 27$, when it is
still realistic to test each value of $y$, we did this. For
greater values of $n$ we tested only the values of $y$ 
satisfying Eq. (\ref{8}) and the values of $y_1$ in some vicinities 
of those $y$ (see below).

Let us first describe the results for $n\leq 27$.          
As follows from Eq. (\ref{16}), the contribution of each $y$ to the
total probability is characterized by the quantity
\begin{equation}
RP(y) = |\frac{1}{A} 
\sum_{j=0}^{j=A-1} g(x(0) + jr, y, n) | ^2              
\label{20}
\end{equation} 
where RP stands for relative probability.
For a given $n$ we chose at random the values of $x(0)$ and
$r$ such that $x(0) < r$ and $r <2^{n/2}$. Then we computed
$RP(y)$ for each $y$. We set some threshold, say 0.05, and
looked for the $y$ passing over that threshold, i.e. for such 
values of $y$ that $RP(y)$ was greater than the threshold. 
Then we computed the continued fraction for $y/N$ and
tested whether it is equal to $k/r$ where $r$ is the period
and $k$ is one of the numbers $0, 1, 2,..r-1$. The result is
that in about 100 computations only the values of $y$ 
satisfying Eq. (\ref{8})
and in some cases $y_1$ such that $|y_1-y|=1$ passed over the
threshold. Moreover, all such values of $y$ passed over the
threshold 0.05. When there were two values, $y$ and $y_1$, 
passing over the
threshold and approximated by $k/r$ then typically RP(y) was
considerably greater than $RP(y_1)$ for the $y$ satisfying Eq. 
(\ref{8}). However we have found several cases when $RP(y_1)$ 
was greater. 

For example, for $n =25,\, x(0) =85,\, r = 713$ both,
$y_1= 23906944$ and $y=23906945$ pass over the threshold and
are approximated by $508/713$. The result is 
$RP(y_1)=0.120148$ and $RP(y)=0.118273$ but it
is the second value which satisfies Eq. (\ref{8}). 

For $n=26,\, x(0) = 211,\, r = 975$ both, $y_1 = 1996058$ and
$y=1996059$ are approximated by $29/975$; we have
$RP(y_1) = 0.106606$ and $RP(y) = 0.0898572$ but
again it is the second value which satisfies Eq. (\ref{8}).          

For $n=27,\, x(0)=163,\, r = 674$ both, $y_1=3186177$ and $y=3186178$
are approximated by $8/337 = 16/674$; $RP(y_1)=0.146263$, 
$RP(y)=0.143943$; again only
the second value satisfies Eq. (\ref{8}).

In all the three cases the quantity $|y_1/N-k/r|$ only 
slightly exceeds $1/2N$.  

For $n>27$ testing each value of $y$ becomes unrealistic and
therefore we should decide what our main priorities are.
What is the main characteristic of the algorithm?
As we already discussed, the probability
to successfully extract the value of $k/r$ from the measurement
outcome depends on $r$ and in favorable
cases can be 1. However such cases are not typical. We should
ask ourselves whether there exists a {\it minimum} probability
of success, such that for all values of $r$ and $x(0)$ the 
probability of success is always {\it greater} than the
minimum probability.

The results for $n\leq 27$ give strong evidence that only values
of $y$ satisfying Eq. (\ref{8}) and possibly some close values 
can essentially contribute to the probability of success.
For this reason we adopt the following approach in the general
case. For given values of $n, x(0)$ and $r$ we test only the
values $[Nk/r]-1, [Nk/r], [Nk/r]+1, [Nk/r]+2$ for $k=1,...r-1$
(the case $k=0$ is obviously trivial) and compute only the
contribution of these values to the probability of success $Pr$.
These quantities represent four numbers in the vicinity of some
$y$ satisfying Eq. (\ref{8}). We denote $Pr(y)$ the {\it relative}
contribution of all these numbers to the probability of success,
i.e. the sum of the quantities $RP(y)$ for those $y$. Let $MinPr(y)$
be the minimum value of $Pr(y)$ for a given run. We also use 
$Pr_{min}$ to denote the minimum value of the total 
probability of success $Pr$ in all our runs for a given $n$. 

The table of the quantities $Pr_{min}$ for $20\leq n \leq 34$
is given below.
  
Our observation is that for odd values of $r$ the probability is
usually very close to $Pr_{min}$, for the values of $r$ divisible by
2 (i.e. for even numbers) it is greater and increases for the 
values of $r$ divisible by 4, 8 etc. This is natural in view of the 
above discussion. For this reason, for $32\leq n \leq 34$ where we
carried out only a few runs, we tested only odd values of $r$. For
such values of $r$ we ran the program on the Windows 2000 machine
equipped with two processors running at 1 GHZ each. The availability
of two processors makes it reasonable to implement the program as a
two-threaded application. Then for $n=32$ it typically takes 4
hours to run each test with a given choice of $x(0)$ and $r$.
For $n=33$ this time becomes 9 hours and for $n=34$ - 20 hours.
We ran four cases for $n=32$, three cases for $n=33$ and two cases
for $n=34$. For each $n$ the results for $Pr$ are very close to each
other and the minimum values of $Pr(y)$ also do not differ 
significantly.

\begin{sloppypar}
For $n=32\,\, (N=4294967296)$ the results are $Pr=0.195057$, 
$MinPr(y)=0.103743$ for
$x(0)=863,\, r =11337$; $Pr=0.195051$, $MinPr(y)=0.119318$ for 
$x(0)=9774,\, r = 22239$;
$Pr=0.195057$, $MinPr(y)=0.120364$ for $x(0)=17867,\, r = 21229$ 
and $Pr=0.195049$, $MinPr(y)=0.103555$ for $x(0)=13559,\, r = 33225$. 

For $n=33\,\, (N=8589934592)$ the results are $Pr=0.185207$,
$MinPr(y)=0.114707$ for $x(0)=17226,\,r=39041$; 
$Pr=0.18524$, $MinPr(y)=0.0967657$ for $x(0)=9244,\,r=18267$
and $Pr=0.185205$, $MinPr(y)=0.0969796$ for 
$x(0)=21533,\,r=27663$.

For $n=34\,\,(N=17179869184)$ the results are $Pr=0.175864$,
$MinPr(y)=0.114707$ for $x(0)=9244,\,r=54337$ and $Pr=0.174863$,
$MinPr(y)=0.103516$ for $x(0)=26700,\,r=36989$.  
\end{sloppypar}

The results confirm our observation for smaller values
of $n$ that when there are no special favorable circumstances, the
value of $Pr$ is almost universal, i.e. practically does not depend
on $x(0)$ and $r$. This shows that our algorithm is very stable,
and $Pr_{min}$ is probably a universal function of $n$. 
The quantities $Pr_{min}$ at different values of $n$ are shown
in Table 1.

\begin{table}
\begin{center}
\caption{{\bf{The quantities $Pr_{min}$ at different values of
$n$ (see text).}}}
\begin{tabular}{|c|cccccc|} \hline
$n$ & 20 & 21 & 22 & 23 & 24 & 25 \\ 
$Pr_{min}$& 0.3630 & 0.3450 & 0.3270 & 0.3108 & 0.2951 & 0.2802 \\\hline
$n$ & 26 &27 & 28 & 29 & 30 & 31 \\
$Pr_{min}$ & 0.2661 & 0.2527 &0.2399 & 0.2278 & 0.2163 & 0.2054 \\\hline
$n$ & 32 &33 &34 & & &  \\
$Pr_{min}$ & 0.1950 &0.1852 &0.1759 & & &\\\hline
\end{tabular}
\end{center}
\end{table}

The data for $MinPr(y)$ are more irregular. In general these values
decrease with the increase of $n$ but are rather sensitive to the
choice of $x(0)$ and $r$. The minimum value of $MinPr(y)$ for all
our runs is equal to $0.0967657$. It was observed for $n=33$ (see 
above), not $n=34$ what might seem to be rather strange and may be
an indication that there exists a minimum of this quantity which
is not equal to zero when $n\rightarrow\infty$. However the
existing amount of data is obviously insufficient for drawing
such a conclusion.

We did not succeed in finding a simple function describing the
data in Table 1. If one tries to approximate the data as
$Pr_{min}(n)=Const/n^c$ then the value of $c$ for $n\in [20,34]$
is in the range $[1.35,1.7]$. 

In the preceding section we have also proposed a modification of
the AQFT (see Eqs. (\ref{AFT1}-\ref{AFT3})). In Table 2 we display
the results of computations of the quantity $Pr_{min}$ in the first
"post integral" approximation corresponding to $m=3$. In this case
decomposition coefficients can take the values of $exp(i\pi l/4)$
$(l=0,1,...7)$ and the problem is no longer formulated only in 
terms of integers. The results show that the probability of period
finding is significantly amplified. Moreover, the results for  
$MinPr(y)$ become much more stable and in all our computations
this quantity was rather close to $Pr_{min}$. The minimum value
of $MinPr(y)$ in our computations is $0.556$ for $n=31$. This does
not improve the estimation $4/\pi^2\approx 0.405$ of the 
minimum relative probability in Shor's algorithm because, as
explained above, $Pr(y)$ represents the contribution of four
values of $y_1$ in the vicinity of $y$ satisfying Eq. (\ref{8}).  
    
\begin{table}
\begin{center}
\caption{{\bf{The quantities $Pr_{min}$ for our modification of
the AQFT at $m=3$ (see text).}}}
\begin{tabular}{|c|cccccc|} \hline
$n$ & 20 & 21 & 22 & 23 & 24 & 25 \\ 
$Pr_{min}$& 0.7568 & 0.7472 & 0.7375 & 0.7282 & 0.7188 & 0.7096 \\\hline
$n$ & 26 &27 & 28 & 29 & 30 & 31 \\
$Pr_{min}$ & 0.7006 & 0.6916 &0.6827 & 0.6740 & 0.6654 & 0.6569 \\\hline
\end{tabular}
\end{center}
\end{table}

It is clear at a glance that the data in Table 2 have a much slower
fall off with the increase of $n$ than those in Table 1. 
If the data are approximated as $Pr_{min}(n)=Const/n^c$ then the 
value of $c$ for $n\in [20,31]$ is in the range $[0.25,0.38]$ i.e.
much better than in the pessimistic estimate (\ref{10}) for the
conventinal AFQT.

\section{Discussion}

In this paper we have proposed a quantum algorithm for factoring
which involves only a finite number of integers. There are strong
numerical indications that the minimum probability to extract the
correct value of $k/r$, where $r$ is the period and $k$ is one of the
numbers $0,1,...r-1$, is a universal function of the number of
qubits $n$ in question. Table 1 in the preceding section
displays minimum probabilities in the range $20\leq n\leq 34$.
where $n=log\, N$ and $N$ is a number used for factorizing a big
number ${\bf N}$. As noted above, the adopted strategy is to choose
$N = c{\bf N}^2$ where $c>1$ is a small number (say in the range
[2,5]). Therefore for big values of ${\bf N}$, $log\,N$ is
proportional to $log\,{\bf N}$.

The numbers in Table 1 are less than the lower bound 
$4/\pi^2\approx 0.405$ for Shor's algorithm and decrease with the 
increase of $n$. Therefore a greater number of repetitions will 
be required to ensure the success. On the other hand, for quantum 
computer implementing our algorithm, the corresponding quantum
circuit has a smaller complexity ($O(n)$ instead of $O(n^2)$),
a much less amount of resources
is required and, since the algorithm involves only integers, its
(in)sensitivity to errors is expected to have the same order of
magnitude than that for classical computer. Classical computer
operating with $n$ bits has $N=2^n$ states while in the version
of quantum computer implementing our factoring algorithm 
with $n$ qubits, the number of states does not exceed $4N$. 
This is a consequence of the fact that the dimension of the Hilbert
space for an $n$-qubit system is equal to $N$ and we need only
linear combinations of basis elements with the coefficients
$1,i,-1,-i$.   

We have also proposed a modification of Coppersmith' Approximate
Quantum Fourier Transform (AQFT). The results in Table 2 show 
that the probability of period finding is signifacantly amplified 
already in the first post integral approximation and the fall off 
with the increase of $n$ is much slower. However this approximation 
no longer can be formulated only in terms of integers. Quantum
computer operating with $n$ qubits in this approximation will
require $8N$ states because now the coefficients can take the
values of $exp(i\pi l/4)$ $(l=0,1,...7)$. In this case it will be 
also necessary to determine a required accuracy for $\sqrt{2}$.
In general it is clear that each next approximation will require
a greater amount of resources and will have a greater 
error-sensitivity.   

At early stages of development of quantum computer our
integral version of Shor's algorithm should be quite sufficient
but for very big numbers one should look for better 
approximations. If one adopts a conventional approach then the
above results give grounds to believe that by using our
modification of the AQFT it will be possible to reduce the 
number of required approximations. At the same time,
it is of indubitable interest to investigate whether there exists 
a quantum factoring algorithm which involves only 
integers and guarantees that the probability of period finding is 
asymptotically constant.

\begin{sloppypar}
{\it Acknowledgements:} The author is grateful to A.A. Makarov
for valuable discussions.
\end{sloppypar}

\end{document}